\documentclass[sigconf]{acmart}

\usepackage[utf8]{inputenc}

\usepackage{enumitem}

\usepackage[multiple]{footmisc}

\newcommand{\resume}[1]{r\'{e}sum\'{e}#1}
\newcommand{\Resume}[1]{R\'{e}sum\'{e}#1}

\definecolor{aqua}{cmyk}{0.91, 0, 0.09, 0.36}

\AtBeginDocument{%
  \providecommand\BibTeX{{%
    \normalfont B\kern-0.5em{\scshape i\kern-0.25em b}\kern-0.8em\TeX}}}
\setcopyright{none}
\renewcommand\footnotetextcopyrightpermission[1]{} 
\begin{document}

\title{
Quantifying the Impact of Human Capital, Job History, and Language Factors on Job Seniority with a Large-scale Analysis of Resumes.
}

\author{Austin P Wright}
\affiliation{
    \institution{Georgia Institute of Technology}
    \city{Atlanta}
    \state{Georgia}
    \country{USA}
}

\author{Caleb Ziems}
\affiliation{
    \institution{Georgia Institute of Technology}
    \city{Atlanta}
    \state{Georgia}
    \country{USA}
}

\author{Haekyu Park }
\affiliation{
    \institution{Georgia Institute of Technology}
    \city{Atlanta}
    \state{Georgia}
    \country{USA}
}

\author{Jon Saad-Falcon}
\affiliation{
    \institution{Georgia Institute of Technology}
    \city{Atlanta}
    \state{Georgia}
    \country{USA}
}

\author{Duen Horng Chau}
\affiliation{
    \institution{Georgia Institute of Technology}
    \city{Atlanta}
    \state{Georgia}
    \country{USA}
}

\author{Diyi Yang}
\affiliation{
    \institution{Georgia Institute of Technology}
    \city{Atlanta}
    \state{Georgia}
    \country{USA}
}

\author{Maria Tomprou}
\affiliation{
    \institution{Carnegie Mellon University}
    \city{Pittsburgh}
    \state{Pennsylvania}
    \country{USA}
}

\renewcommand{\shortauthors}{Wright, et al.}

\begin{abstract}

As job markets worldwide have become more competitive and applicant selection criteria have become more opaque, and different (and sometimes contradictory) information and advice is available for job seekers wishing to progress in their careers, it has never been more difficult to determine which factors in a \resume{} most effectively help career progression. In this work we present a novel, large scale dataset of over half a million \resume{s} with preliminary analysis to begin to answer empirically which factors help or hurt people wishing to transition to more senior roles as they progress in their career. We find that previous experience forms the most important factor, outweighing other aspects of human capital, and find which language factors in a \resume{} have significant effects. This lays the groundwork for future inquiry in career trajectories using large scale data analysis and natural language processing techniques. 
\end{abstract}

\maketitle
\thispagestyle{empty}
\pagestyle{plain} 

\section{Introduction}

Given the increasingly competitive global job market, job seekers are interested in finding the most effective ways to advance their careers with investment in education, skills training, and \resume{} fine tuning. In this work we present a first-of-its-kind, large-scale analysis of a \resume{} corpus with 641,170 resumes. We operationalize a large set of factors by utilizing natural language processing (NLP) and data science methods to computationally model how human capital factors (e.g., education, skills), \resume{} presentation language, and previous experience, jointly predict career transitions into senior positions in order to address the following open research questions: 

\begin{enumerate}[label=\textbf{RQ\arabic*},itemsep=1mm, topsep=1mm,parsep=1mm, leftmargin=10mm]

\item How do seniority transitions occur across the workforce over time within people's careers?

\item How does human capital --- conceptualized as the accumulation skills, and education --- affect seniority transitions?

\item How does employment history including gaps between jobs affect seniority transitions?

\item How much of an impact does \resume{} language, style, and presentation have on seniority transitions? 

\item How do these effects apply across sectors of occupations/jobs? 

\end{enumerate}

\section{Related Work}

\subsection{Career Transitions}
Career transitions and success have been of great interest in economics \cite{jovanovic1984matching,feldman2007careers} and psychology \cite{judge1999big,ng2005predictors}. 
In computer science, there is growing interest in predicting career transitions and job tenure at the individual level using statistical methods like survival analysis \cite{li2017prospecting} and generalized linear models \cite{xu2014modeling}, as well as neural networks \cite{li2017nemo,meng2019hierarchical}. 
Individuals can be modeled as a sequence of their professional positions, where each position node is indicated by the job title, company, industry, time duration, and keyword summary of the job. \cite{xu2014modeling} compute node-level similarity using GLMs and compute user-level similarity by aligning users' node-level similarity time series.
Another body of work considers talent flows in a job transition networks where organizations are represented as nodes and the flow of employees between these organizations is represented by a set of weighted edges \cite{xu2016talent,xu2018dynamic}. 
Safavi et al. \cite{safavi2018career} analyze the career trajectories of computer science PhD students, identifying hubs and authorities in the career transition networks, which include movements between employers and also sectors (academia, government, industry). 
Xu et al. \cite{xu2018dynamic} used stock market prices to predict changes in the job transition network across time.
There is also significant amount of research studying how to develop job recommendation systems \cite{wang2013time,kenthapadi2017personalized,zhang2014research,mishra2016bottom,shalaby2017help}.

\subsection{Resume Analysis}
\label{subsec:resume_quality}

In order to study careers on a large scale, \resume{s} form an accessible and comprehensive source of data of job history, human capital, and language framing. 
Zhang and Wang \cite{zhang2018resumevis} proposed a \resume{} visualization system, Resumevis, which they use to summarize patterns of career development, compare career trajectories across \resume{s}, find similar \resume{s}, and discover career evolving patterns across time and organizations in the aggregate. 
Other visualization work has developed standard ways to extract information from CV's and more easily compare across formatting \cite{filipov2019cv3}.
Additionally there are many resources for \resume{} writers for proposed best practises. 
High-quality \resume{s} are free of spelling or grammatical mistakes \cite{risavy2017resume}, and the best \resume{s} are typically short and concise \cite{helwig1985corporate}.
Furthermore, there is an empirical advantage to using \textit{strong active verbs} like ``built,'' ``presented,'' and ``manufactured'' instead of vague generalizations or platitudes like ``amplified company impact'' \cite{gross201712,oliphant1982reactions,stephens1979getting}. 
Power and agency verb frames can also connote that the applicant is ``decisive'' and ``capable of pushing forward their own storyline'' \cite{sap2017connotation}. 
On the other hand, hedge words indicate uncertainty and can weaken the applicant's rhetorical persuasiveness \cite{hyland2018metadiscourse}. 
Similarly, subjective language (\textit{considerably, courageously, crazy, delightful}) can weaken a \resume{} \cite{riloff2003learning}. 
Professionals expect accomplishments to be linked to concrete outcomes and quantified when possible \cite{gross201712}. 
Finally, applicants are expected to avoid using personal pronouns \cite{horn1988employers}, passive voice \cite{gross201712}, or any overly emotional language \cite{weaver2017predicting}.

\section{Data}

We scraped 635,929 public \resume{s} posted in August 2017 to an online \resume{} storage and hiring service. 

\begin{figure}
    \centering
    \includegraphics[width=0.85\columnwidth]{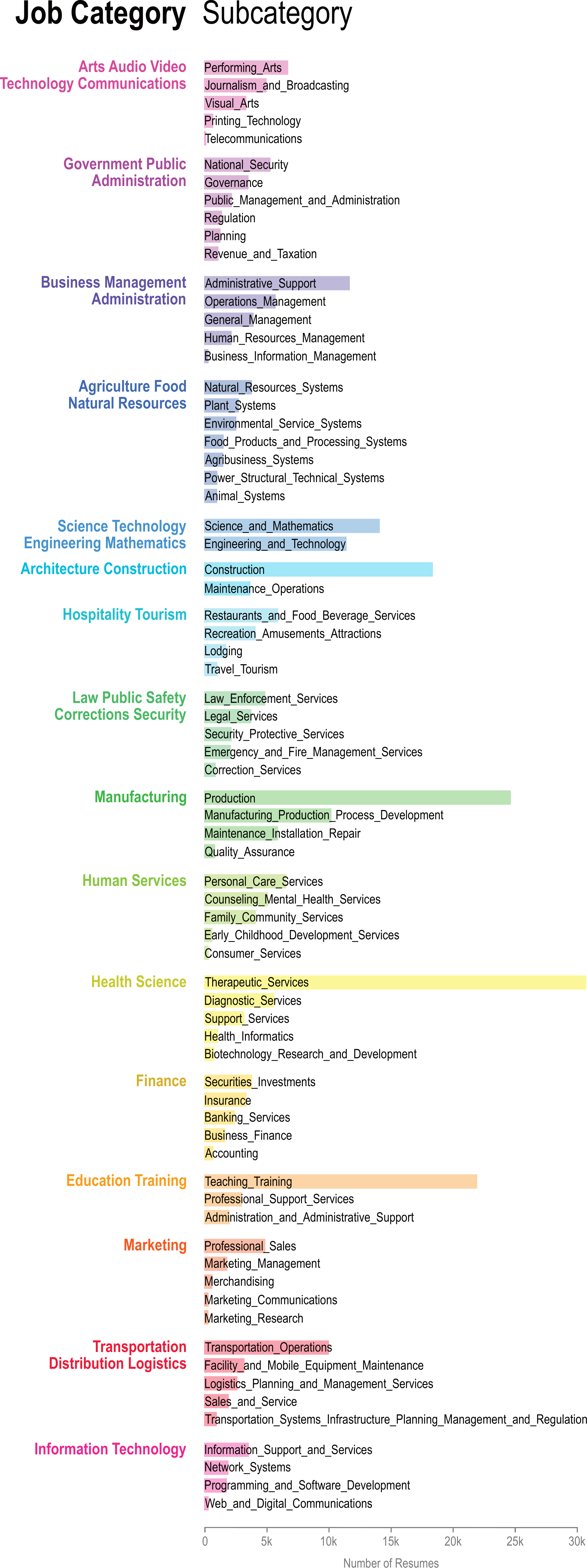}
    \caption{
        Job Category Distribution.
        Our \resume{} data contains 16 main categories, and each category (left column) has several subcategories (right column).
        Bars under subcategories show the number of \resume{s} of the corresponding subcategory.
    }
    \label{fig:job_cat_dist}
\end{figure}

We used the Standard Occupational Classification (SOC) System from Bureau of Labor Statistics (\url{https://www.bls.gov/soc/}) to structure resumes. The \resume{s} span 16 occupational sectors specified in the SOC: 
Manufacturing, Health Science, Education, STEM, Business, Architecture, Transportation, Human Services, Arts \& Communication, Government, Food \& Agriculture, Public Safety, Tourism, Finance, Marketing, and Information Technology (IT).
For each occupation sector, we used the job titles defined within the sector as query terms for \resume{s}.
Each sector is well-represented in our dataset, with 15,000 samples in IT, the least prevalent sector. 
The distribution of job categories can be seen in Figure \ref{fig:job_cat_dist}.

\begin{figure}[t]
    \centering
    \includegraphics[width=0.85\columnwidth]{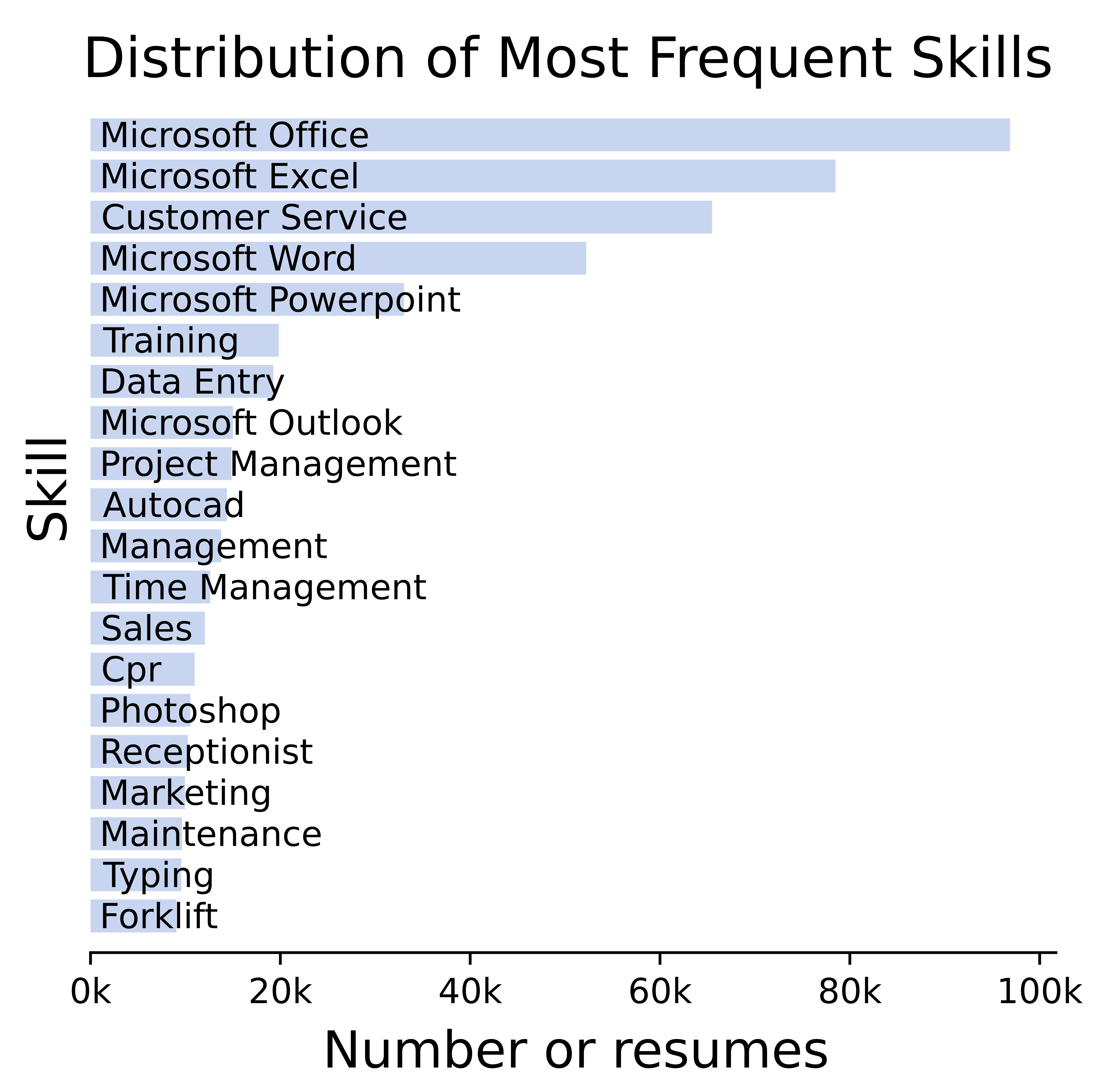}
    \caption{
        Distribution of skills most frequently mentioned in our \resume{} data.
    }
    \label{fig:skill_dist}
\end{figure}

\begin{figure}[t]
    \centering
    \includegraphics[width=0.85\columnwidth]{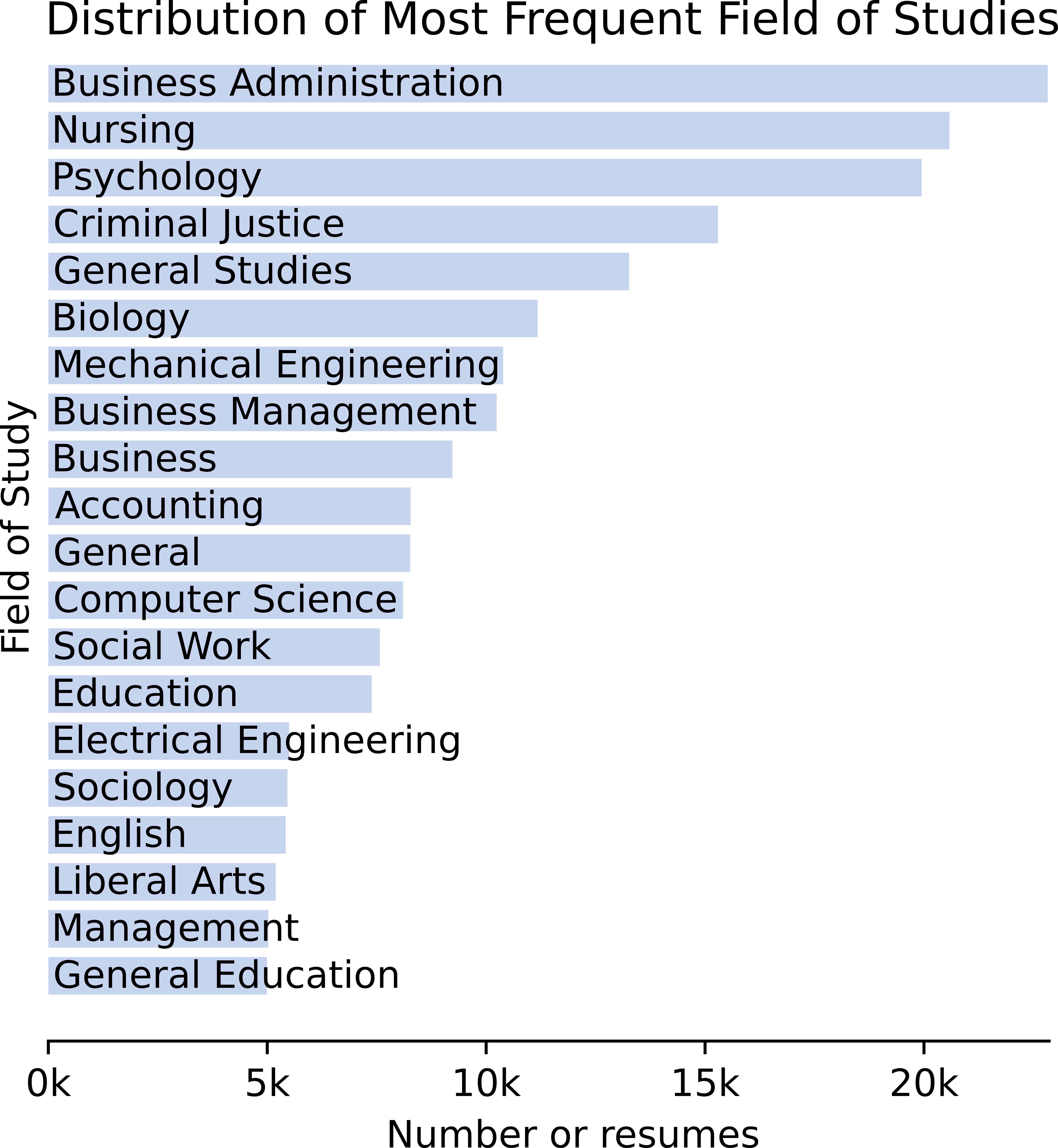}
    \caption{
        Distribution of fields of studies most frequently mentioned in our \resume{} data.
    }
    \label{fig:field_of_study_dist}
\end{figure}

\begin{figure}[t]
    \centering
    \includegraphics[width=0.85\columnwidth]{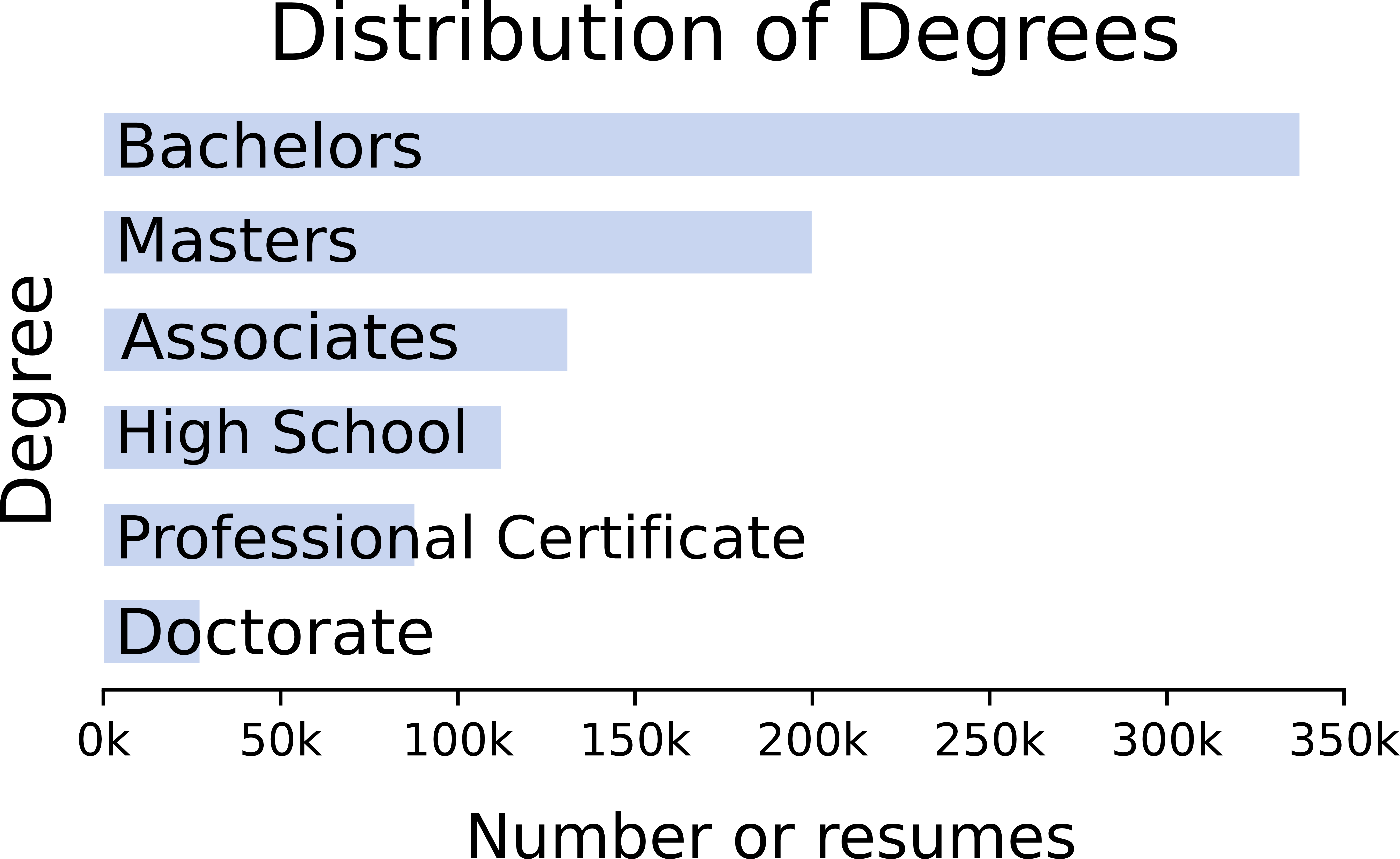}
    \caption{
        Distribution of degrees
    }
    \label{fig:degree_dist}
\end{figure}

\begin{figure}[t]
    \centering
    \includegraphics[width=0.85\columnwidth]{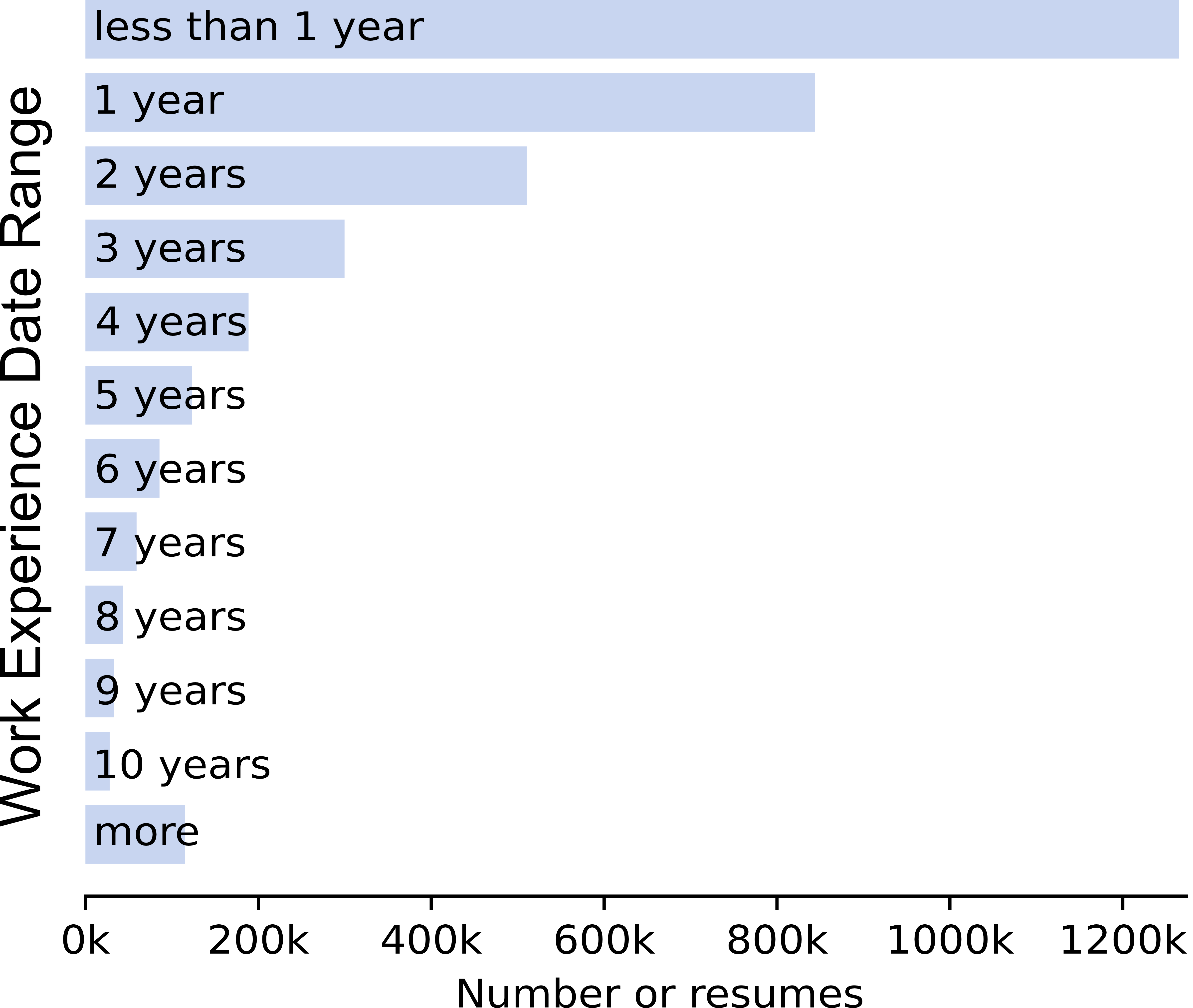}
    \caption{Distribution of work experience date range}
    \label{fig:exp_range_dist}
\end{figure}

As a pre-processing step, we removed duplicate \resume{s} and converted all documents to a regular JSON format with the following fields:

\begin{enumerate}
    \item \textbf{Summary}: This optional field corresponds to the ``objective'' statement that some \resume{} authors chose to include.
    
    \item \textbf{Education}: This field represents the educational levels achieved by the individual, including the degree (as shown in Figure \ref{fig:degree_dist}), field of study (as shown in Figure \ref{fig:field_of_study_dist}), and date range.

    \item \textbf{Work Experience}: This field includes a the work experience of the individual. It is a list  of job entries each with a date range (Figure \ref{fig:exp_range_dist}), description, and title.
    
    \item \textbf{Skills}: This field is a list of reported skills accompanied by a value representing months of experience in that skill. Figure \ref{fig:skill_dist} shows the distribution of 20 most frequent skills.

    \item \textbf{Awards}: This field represents the awards earned by the individual with names and descriptions.

    \item \textbf{Patents}: This field represents the patents held by the individual, with names and descriptions. 
    
    \item \textbf{Certifications}: This field holds the certifications accumulated by the individual with names and descriptions. 
    
    \item \textbf{Publications}: This field represents the publications of the individual listed in their profile with names and descriptions. 
\end{enumerate}

\section{Measures}
\subsection{Extracting Education}
In our dataset, education is self reported as a list of attained education levels, with descriptions of the `field' of the degree, the `dateRange', and the actual `degree' all self reported strings.
In order to categorize these into standard education levels, we defined the most common (American) degree/education types:
\begin{enumerate}
    \item High School
    \item Associates
    \item Bachelors
    \item Masters
    \item Doctorate
    \item Professional Certificate
\end{enumerate}

For each of degree types, we enumerated common abbreviations and names referring to each type.
We then categorized each \resume{} by the most recent education entry, taking the `degree' field and searching for matches of these known terms. The terms used for each education type are:

\begin{enumerate}
    \item High School \\ `high', `hs', `h.s.', `ged', `g.e.d.'
    \item Associates \\  `aa', `a.a.', `associate', `associates', `assoc', `a.a', `aa.'
    \item Bachelors \\  `bachelor', `bachelors', `bachelor\'s', `ba', `bs', `b.a.', `b.s.', `b.a', `b.s', `bfa', `b.f.a.'
    \item Masters \\ `master', `masters', `ms', `ma', `mfa', `m.s.', `m.a.', `m.f.a.', `msc', `m.a', `ma.', `m.s', `ms.', `msc.', `m.b.a', `mba', `mph', `m.p.h.', `mpa', `m.p.a.', `m.'
    \item Doctorate \\ `phd', `phd.', `ph.d', `ph.d.', `doctorate', `doctor', `jd', `j.d.', `md', `m.d.', `d.'
    \item Professional Certificate \\ `cert', `certificate', `certification', `vocation', `vocational'
\end{enumerate}

In addition, we categorized \resume{s} without education descriptions (or with empty education descriptions) with the category `None'. 
These categories together account for 93\% of all resumes, 
with the remaining \resume{s} not including any of these descriptions, and we put them in the `Other' category.
These ``Other'' \resume{s} often contain ambiguous education descriptions, 
e.g., ``college'', 
``39 credits earned'', 
or ``visual communications''.

\subsection{Job History}
For each \resume{}, we extract information regarding their job history. 
Each \resume{} includes a list of job entries, each containing information regarding the date range of the job. From this list, we extract multiple pieces of information:

\begin{enumerate}

    \item The total number of job entries. 

    \item The total amount of time working. This is done by taking each job entry and calculating the time worked at that job using the start and end dates. Then this value is summed over every job entry. In the case where multiple jobs are listed overlapping in time, that overlapped time is counted for each job separately and are both part of the total number. 

    \item The size of the largest gap in employment on the resume. 
    A gap is defined as the amount of time (number of months) between the end date of a job and the start date of the next job.
    Thus, the largest gap refers to the longest such time interval. 

    This metric is potentially useful for  understanding the impact of large gaps caused by maternity leave, medical issues, incarceration, or other causes of gaps which are often pointed out as negative factors on \resume{s}.

\end{enumerate}

\subsection{Language Factors}
We operationalized professionally-attested writing strategies as explicit prescriptive measures of \resume{} quality, 
summarized in \autoref{tab:resume_tips}. 
To begin, we compiled a collection of relevant lexicons and phrase sets from public career advice pages on the web. 
We then set out to implement analysis of each \resume{} by looking at the job descriptions and summary statement (if present). 

We built phrase sets of \textsc{buzzwords}
\footnote{\url{https://en.wikipedia.org/wiki/Buzzword}}\footnote{\url{https://sba.thehartford.com/business-management/delete-these-buzzwords/}}\footnote{\url{https://github.com/words/buzzwords}}, 
\textsc{graded quantifiers} 
and \textsc{qualifiers}. 
Additionally, 
we considered relevant \textsc{hedge} \cite{hyland2018metadiscourse}
 
lexicons from previous works on text and discourse analysis. 

Finally, we applied a multi-step NLP pipeline to operationalize our prescriptive \resume{} quality features. 
On each sentence in the \resume{}, we ran the standard \texttt{spaCy} language processing pipeline, 
which includes tokenization (with stemming), part of speech (POS) tagging, 
dependency parsing, 
and named-entity recognition (NER). 
For each lexicon, we counted matches at the stemmed token level, and for each phrase set, we used a string matching procedure to capture multi-word phrases. 

Separately, we found real-valued sentiment scores using VADER \cite{gilbert2014vader}.

\begin{table*}[ht]
    \centering
    \resizebox{1.0\textwidth}{!}{%
    \begin{tabular}{p{60mm}p{50mm}p{85mm}}
        \toprule
        \textbf{Measure} & \textbf{Approach} & \textbf{Examples} \\
        \midrule
        
        Use active voice (not passive) \cite{gross201712}
        & flag \texttt{nsubjpass} dependencies 
        & \textbf{passive:} ``Sales \textit{were increased} 30\% after my system was deployed'' \newline \textbf{active:} ``My system \textit{increased} sales by 30\%''
        \\ \hline

        Use exact quantities (not graded quantifiers) \cite{gross201712} 
        & use \textsc{graded quantifiers} phrase set \newline flag numeric entity types (SpaCy)
        & \textbf{graded quantifier:} ``My approach improved sales \textit{a lot}'' \newline \textbf{exact quantity:} ``My approach improved sales by 30\% \\ \hline
        
        Avoid buzzwords 
        & use \textsc{buzzwords} phrase set 
        & \textbf{buzzwords:} \textit{team player, synergy, next generation, moving forward} \\ \hline 
        
        Avoid semantically vacuous words and symbols \cite{gross201712,oliphant1982reactions,stephens1979getting} 
        & flag \texttt{fixed} dependencies \newline flag \texttt{discourse} dependencies \newline flag \texttt{expl} dependencies \newline flag \texttt{vocative} dependencies 
        & 
        \textbf{fixed MWE:} 
        \textit{as well as, in addition to, because of} \newline 
        \textbf{discourse:} 
        \textit{like, well, oh, uh, :)} 
        \newline 
        \textbf{expl:} 
        ``\textit{There} were 60 people in attendance'' \newline 
        \textbf{no expl:} 
        ``60 people attended'' 
        \newline 
        \textbf{vocative:} 
        ``\textit{Guys,} the software really works perfectly'' 
        \newline 
        \textbf{no vocative:} ``The software works perfectly'' 
        \\ \hline
        
        Avoid qualifiers and other hedges \cite{hyland2018metadiscourse} & use \textsc{hedges} lexicon \cite{hyland2018metadiscourse} \newline use \textsc{qualifiers} phrase set & \textbf{hedges:} \textit{approximately, generally, often, seems, tends to, indicates} \newline \textbf{qualifiers:} \textit{actually, basically, just, kind of, really, sort of}\\ \hline
        
        Avoid overly emotional or subjective language \newline (and especially avoid negativity) \cite{weaver2017predicting} & use VADER sentiment \cite{gilbert2014vader} \newline use weak and strong \textsc{subjectivity} lexicons \cite{riloff2003learning} & \textbf{strongly subjective:} \textit{considerably, courageously, crazy, delightfully}
        \textbf{strongly negative:} I \textit{abhor} public speaking... Technology is so \textit{annoying}
        \\ \hline
        
        Remove all personal pronouns \cite{horn1988employers} & POS tagging & \textbf{first-person personal pronoun}: \textit{I} implemented a \texttt{seq2seq} model. \newline \textbf{third-person personal pronoun}: \textit{We} implemented a \texttt{seq2seq} model. \newline \textbf{no personal pronoun}: Implemented a \texttt{seq2seq} model.\\ \hline
        
        Use past-tense verbs & POS tagging & \textbf{past:} ``Operated the lathe'' \newline \textbf{present continuous:} ``Operates the lathe'' \\ \hline
        
        Remove all ``be'' verbs & match lemmas & \textbf{be verb:} ``I \textit{was} responsible for operating the lathe'' \newline \textbf{no be verb:} ``Operated the lathe'' \\ \hline

        \bottomrule
    \end{tabular}}
    \caption{Language Factors of \Resume{} Quality}
    \label{tab:resume_tips}
\end{table*}

\subsection{Extracting Senior Positions}

In order to understand how people advance in their careers we defined a measure of job title seniority. Senior jobs are those that represent the higher levels within the management hierarchy of an organization, thus representing managers, executives, and other jobs in charge of managing others. We define Senior positions with a list of representative terms often used in such job titles. For example an archetypal `Senior' position would be a management or executive level position, and thus such a title may include terms like `lead', `head' or `manager'. 
Thus we define a set of paradigmatic Senior title words as: `senior', `chief', `lead', `head', `president', `manager', `director', `supervisor', 'superintendent', `ceo', `coordinator', `principal', `founder', `partner', `sr', `cfo', `cio'.
The expectation is that any senior job title would include some terms within or similar to the specific words mentioned above.
As a way of measuring the similarity between job title variants, we use word mover distance \cite{kusner2015word}, a metric designed to measure the similarity between word embeddings that capture the syntactic or semantic meaning of sentences. 
By using this metric to gauge how similar two different job titles are, we can match job titles that do not use identical words but effectively describe the same roles. This analysis results in a classification of Senior or non-Senior for every job history entry in each \resume{}, with 24\% of resumes being currently in senior positions and thus 76\% of positions being currently non-senior.

\section{Analysis}

Given our large and feature-rich dataset, we set forth to analyze how each of the factors we extracted from each \resume{} affects career trajectories. 
In particular, we looked at what helps enable people to transition into more senior roles as they progress in their career (RQ1). We did this by performing a linear logistic regression analysis (with l1 regularization) with a dependent variable of the most recent job seniority categorization on each \resume{}.

As independent variables, we included the seniority level of each person's second most recent job entry, to see the comparative effect of already having a senior position, and thus the likelihood of maintaining such a position or transitioning from a non-senior position into a senior position. 
We included the total number of jobs held
and the total cumulative time of work experience
to see the effect of experience.
In addition, we include human capital factors (RQ2) of the highest level of education achieved, the number of skills, awards, publications, and patents as an independent variables. 
Furthermore, we include the size of the largest employment gap in each \resume{} to measure the effect of such gaps in upward mobility (RQ3). 
Finally, we also include the language factors extracted from each \resume{} as independent variables, allowing us to compare how impactful these actionable strategies are in comparison to other factors (RQ4) in transitioning to senior roles. 

We repeated this regression analysis for \resume{s} in each high level Job Sector in order to see how these factors differed between sectors (RQ5).

\subsection{Seniority Transitions}
Given that the information used in the regression formulation is an incomplete picture of each career, we expect that any analysis is bound to be incomplete as well. 
In order to understand what portion of the real world effect our model is characterizing we evaluated our logistic regression first against a naïve baseline that always guesses 
the more likely option of non-senior
as a lower bound of accuracy to assess our model's improvement over the baseline. 
Table \ref{tab:regression overview} shows the evaluation of the accuracy and F1 score of the regression compared to baseline (F1 score being more relevant due to the unbalanced nature of more non-senior positions). 
The regression performs better than baseline across all job sectors, suggesting a measurable gain in information and predictive power  of the features extracted from the dataset. 
Furthermore, we can also see the differential in how easy or difficult each sector is to predict, with 
Business Management Administration having the largest increase, 
which makes sense as our encoding of seniority may be the most relevant in this sector, and 
Human Services having the smallest increase showing, possibly due to the large variety of jobs in that category, and showing the limitations of this technique in other contexts. 
Once it is established that the regression has learned predictive features, we can analyze the parameters of the regression to see the comparative effect of each independent variable on the dependent variable of likelihood to progress into senior positions. 

\begin{table*}[ht]
    \centering
    \resizebox{1.0\textwidth}{!}{%
    \begin{tabular}{l|rrrr}
        \toprule
        \textbf{Job Sector} & \textbf{Regression Accuracy} & \textbf{Baseline Accuracy}& \textbf{Regression F1 Score} & \textbf{Baseline F1 Score}\\
        \midrule
        Agriculture Food Natural Resources &  0.757405  & 0.739286 & 0.583977 & 0.425051 \\
        Architecture  Construction  & 0.815911  & 0.800044  & 0.616268  & 0.444458\\
        Arts  Audio  Video  Technology  Communications  & 0.726572  & 0.709866  & 0.558111  & 0.415159\\
        Business  Management  Administration  & 0.735396  & 0.684728  & 0.660820  & 0.406432\\
        Education  Training  & 0.798285  & 0.794900  & 0.485449  & 0.442866\\
        Finance  & 0.730177  & 0.711849  & 0.561279  & 0.415836\\
        Government  Public  Administration  & 0.714211  & 0.691448  & 0.610605  & 0.408791\\
        Health  Science  & 0.831652  & 0.830961  & 0.498625  & 0.453839\\
        Hospitality  Tourism  & 0.737358  & 0.729057  & 0.540708  & 0.421650\\
        Human  Services  & 0.769675  & 0.768184  & 0.471345  & 0.434448\\
        Information  Technology  & 0.727976  & 0.693597  & 0.643646  & 0.409541\\
        Law  Public  Safety  Corrections  Security  & 0.810607  & 0.810382  & 0.474382  & 0.447630\\
        Manufacturing  & 0.816519  & 0.795754  & 0.614662  & 0.443131\\
        Marketing  & 0.711056  & 0.686147  & 0.581087  & 0.406932\\
        Science  Technology  Engineering  Mathematics  & 0.744574  & 0.693599  & 0.661161  & 0.409541\\
        Transportation  Distribution  Logistics  & 0.785876  & 0.767633  & 0.609997  & 0.434272\\
        \bottomrule
    \end{tabular}}
    \caption{Evaluation of regressions compared to naïve baseline (always guesses the more likely option of \textit{non-senior}), showing a higher accuracy and F1 score across all job sectors.}
    \label{tab:regression overview}
\end{table*}

\subsection{Previous Experience}
The first factor we analyze is the effect of previous job experience on senior transitions. 
We expected that more work experience would help achieve seniority; 
and in alignment with RQ3, we expect that larger employment gaps would have a negative impact. 
Table \ref{tab:job_exp} shows that the most impactful parameter is the total time of job experience, having a significant positive impact on seniority across sectors which agrees with our hypothesis that people with more years of experience are more likely to be promoted to senior positions. 
At the same time, the total number of jobs has little or no significant impact in most industries. 
Finally, we see the effect of employment gaps is, as expected, negative across job sectors. 
However this effect seems to be comparatively small with less overall significance in most sectors. 
The sectors with the biggest impact of gaps are  Manufacturing, Business Management Administration, and Education Training. 

\begin{table*}[ht]
    \centering
    \resizebox{1.0\textwidth}{!}{%
    \begin{tabular}{l|rrr}
        \toprule
        \textbf{Job Sector} & \textbf{Total Job Time} & \textbf{Number of Jobs} & \textbf{Largest Gap Size}\\
        \midrule
        Agriculture Food Natural Resources & 0.282±0.017 & 0.000  & -0.028±0.018 \\
        Architecture Construction & 0.294±0.017 & -0.029±0.017 & -0.019±0.016 \\
        Arts Audio Video Technology Communications & 0.132±0.018 & -0.018±0.018 & -0.036±0.017 \\
        Business Management Administration & 0.297±0.015 & -0.009±0.015 & -0.128±0.015 \\
        Education Training & 0.186±0.013 & 0.000  & -0.118±0.015 \\
        Finance & 0.242±0.018 & 0.000  & -0.067±0.020 \\
        Government Public Administration & 0.162±0.016 & 0.000  & -0.064±0.017 \\
        Health Science & 0.213±0.012 & 0.000  & -0.065±0.013 \\
        Hospitality Tourism & 0.173±0.019 & -0.041±0.020 & -0.055±0.019 \\
        Human Services & 0.193±0.016 & 0.000  & -0.069±0.018 \\
        Information Technology & 0.158±0.032 & 0.000  & -0.056±0.036 \\
        Law Public Safety Corrections Security & 0.190±0.020 & 0.025±0.020 & -0.016±0.019 \\
        Manufacturing & 0.361±0.011 & 0.000  & -0.131±0.013 \\
        Marketing & 0.074±0.022 & 0.000  & -0.074±0.023 \\
        Science Technology Engineering Mathematics & 0.329±0.016 & 0.006±0.015 & -0.056±0.013 \\
        Transportation Distribution Logistics & 0.174±0.017 & -0.002±0.017 & -0.081±0.018 \\
        \bottomrule
    \end{tabular}}
    \caption{Effects of job experience factors on predicting seniority. Total job time shows a significant positive impact across sectors, number of jobs show little effect, and employment gaps have a smaller but significant negative effect. The table shows coefficients in the logistic regression with standard error intervals. Larger positive numbers imply a greater positive impact for senior transitions, while more negative numbers imply a penalty, values near zero imply little effect.}
    \label{tab:job_exp}
\end{table*}

\subsection{Human Capital}
In accordance with RQ2 we looked at the effect of human capital on Seniority. 
Our hypothesis is that higher levels of education as well as skills would have a positive impact on transitions to senior positions. 
Table \ref{tab:human_capital} shows our findings. 
We find that contrary to our hypothesis the effect of human capital is often either very small or we find no significant effect in most sectors. 
In education, we find small penalties for attaining education levels below a Bachelor level (Associates, Certificate, High School, None, and Other), 
and a significant but small positive effect for having a Bachelor's degree, however in most fields there is little or no added effect from further education at the masters or doctorate level. 
The effect of education however does vary widely across sectors, with higher education being  more helpful in Business   Management    Administration and Health Science, 
but less measurable in Arts  Audio Video Technology Communications, Marketing, and Agriculture Food Natural Resources. 
At the same time, skills, publications, awards, and certifications have marginal or negligible effects across all sectors. 
This may be because the context of this analysis is looking at management level seniority which in general may not ask for as many listed technical skills compared to other kinds of advancement in a career on a non-managerial track.

\begin{table*}[ht]
    \centering
    \resizebox{1.0\textwidth}{!}{%
    \begin{tabular}{l|rrrrrrrr}
        \toprule
        \textbf{Job Sector} & 
        \textbf{ Associates } &
        \textbf{ Bachelors } &
        \textbf{ Certificate } &
        \textbf{ Doctorate } &
        \textbf{ HighSchool } &
        \textbf{ Masters } &
        \textbf{ None } &
        \textbf{ Other } \\
        \midrule
        Agriculture Food Natural Resources  & -0.042±0.019 & 0.059±0.019 & -0.030±0.018 & 0.000 & -0.103±0.023 & 0.000 & 0.000 & -0.018±0.018  \\
        Architecture Construction  & 0.000 & 0.138±0.015 & -0.014±0.017 & 0.000 & -0.146±0.021 & 0.000 & 0.000 & -0.041±0.017  \\
        Arts Audio Video Technology Communications  & -0.033±0.019 & 0.019±0.023 & -0.037±0.018 & 0.000 & -0.113±0.022 & 0.009±0.021 & 0.000 & -0.065±0.019  \\
        Business Management Administration  & -0.062±0.016 & 0.134±0.017 & -0.045±0.015 & 0.059±0.013 & -0.099±0.017 & 0.072±0.017 & 0.000 & 0.000  \\
        Education Training  & -0.008±0.014 & 0.022±0.015 & -0.013±0.014 & 0.064±0.013 & -0.042±0.018 & 0.000 & 0.000 & -0.003±0.014 \\
        Finance  & -0.028±0.020 & 0.000 & 0.000 & 0.000 & -0.014±0.021 & 0.057±0.020 & -0.026±0.020 & -0.034±0.020 \\
         Government Public Administration  & -0.035±0.017 & 0.000 & -0.016±0.017 & 0.000 & -0.046±0.018 & 0.054±0.017 & 0.000 & -0.017±0.017  \\
        Health Science  & -0.091±0.018 & 0.120±0.018 & -0.026±0.016 & 0.046±0.014 & -0.057±0.017 & 0.073±0.019 & 0.000 & -0.004±0.015 \\
        Hospitality Tourism  & 0.035±0.019 & 0.110±0.019 & 0.000 & 0.000 & -0.077±0.021 & 0.000 & 0.000 & -0.014±0.020 \\
        Human Services  & 0.063±0.016 & 0.088±0.017 & -0.016±0.017 & 0.019±0.017 & -0.065±0.020 & 0.000 & 0.000 & -0.023±0.018 \\
        Information Technology  & 0.000 & 0.000 & 0.000 & 0.000 & 0.000 & 0.000 & 0.000 & 0.000  \\
        Law Public Safety Corrections Security  & 0.000 & 0.082±0.021 & -0.006±0.021 & 0.000 & 0.000 & 0.046±0.021 & 0.000 & 0.000  \\
        Manufacturing  & -0.047±0.013 & 0.087±0.012 & -0.056±0.013 & 0.016±0.011 & -0.114±0.016 & 0.000 & 0.000 & -0.073±0.013  \\
        Marketing  & 0.000 & 0.025±0.025 & 0.000 & 0.000 & -0.007±0.023 & 0.079±0.025 & 0.000 & -0.040±0.024 \\
        Science Technology Engineering Mathematics  & -0.051±0.014 & 0.000 & -0.002±0.013 & 0.000 & -0.026±0.015 & 0.013±0.013 & 0.000 & -0.025±0.013  \\
        Transportation Distribution Logistics  & -0.004±0.019 & 0.115±0.021 & -0.014±0.019 & 0.000 & -0.121±0.022 & 0.007±0.022 & 0.000 & -0.061±0.020  \\
        \bottomrule
    \end{tabular}}
    \caption{Effects of highest degree earned on predicting seniority showing small positive effects for Bachelor's degrees across sectors. 
    The table shows coefficients from the logistic regression with standard error intervals. Larger positive numbers imply a greater positive impact for senior transitions, while more negative numbers imply a penalty, values near zero imply little effect.}
    \label{tab:education}
\end{table*}

\begin{table*}[ht]
    \centering
    \resizebox{1.0\textwidth}{!}{%
    \begin{tabular}{l|rrrrr}
        \toprule
        \textbf{Job Sector} & 
        \textbf{ Number Skills } &
        \textbf{ Number Awards } &
        \textbf{ Number Certifications } &
        \textbf{ Number Publications } &
        \textbf{ Number Patents } \\
        \midrule
        Agriculture Food Natural Resources & 0.000 & 0.000 & 0.000 & 0.000 & 0.000 \\
Architecture Construction  & 0.002±0.016 & 0.000 & 0.000 & 0.003±0.015 & 0.002±0.016 \\
Arts Audio Video Technology Communications  & -0.027±0.016 & 0.000 & 0.000 & -0.006±0.016 & 0.000 \\
Business Management Administration & -0.010±0.014 & 0.017±0.012 & 0.000 & 0.000 & 0.000 \\
Education Training  & 0.000 & 0.009±0.014 & 0.031±0.014 & 0.000 & 0.000 \\
Finance  & 0.000 & 0.000 & -0.023±0.018 & 0.000 & 0.000 \\
Government Public Administration & 0.000 & 0.000 & 0.031±0.017 & 0.000 & 0.000 \\
Health Science & 0.052±0.012 & 0.038±0.011 & -0.081±0.014 & 0.000 & 0.005±0.010 \\
Hospitality Tourism  & 0.000 & 0.000 & 0.003±0.019 & 0.000 & 0.000 \\
Human Services  & 0.018±0.016 & 0.000 & 0.000 & 0.000 & 0.000 \\
Information Technology  & 0.000 & 0.000 & 0.000 & 0.000 & 0.000 \\
Law Public Safety Corrections Security & 0.000 & 0.000 & 0.000 & 0.001±0.017 & 0.000 \\
Manufacturing & 0.006±0.012 & 0.013±0.011 & 0.000 & 0.002±0.012 & 0.017±0.012 \\
Marketing  & 0.000 & 0.004±0.021 & 0.000 & 0.000 & 0.000 \\
Science Technology Engineering Mathematics & -0.033±0.013 & -0.009±0.014 & 0.040±0.013 & 0.000 & 0.011±0.012 \\
Transportation Distribution Logistics & 0.033±0.016 & 0.000 & -0.068±0.020 & 0.000 & 0.000 \\
        \bottomrule
    \end{tabular}}
    \caption{Effects of human capital factors on predicting seniority showing minimal effects of self reported skills and accomplishments. The table shows coefficients from the logistic regression with standard error intervals. Larger positive numbers imply a greater positive impact for senior transitions, while more negative numbers imply a penalty, values near zero imply little effect.}
    \label{tab:human_capital}
\end{table*}

\subsection{Language Factors}
Finally, we looked at the effect of \resume{} language factors on seniority transitions with results in Tables \ref{tab:lang_factors_senitment}, \ref{tab:lang_factors_phrases}, and \ref{tab:lang_factors_syntax}.  
Our current results 
We found that while experts recommend avoiding buzzwords, the inclusion of buzzwords has a large positive impact (with an effect of nearly the same size as cumulative working time for most sectors) across all job sector; other recommendations have much less impact, which is a surprising result and warrant further investigation (e.g., language factors could have a stronger effect for \resume{s} with less job experience).

\begin{table}[ht]
    \centering
    \resizebox{1.0\columnwidth}{!}{%
    \begin{tabular}{p{40mm}|rr}
        \toprule
        \textbf{Job Sector} & 
        \textbf{ Negative Sentiment } &
        \textbf{ Positive Sentiment } \\
        \midrule
Agriculture Food Natural Resources  & 0.000 & 0.000 \\
Architecture Construction  & -0.012±0.017 & -0.006±0.017 \\
Arts Audio Video Technology Communications  & 0.010±0.016 & 0.000 \\
Business Management Administration  & 0.000 & -0.057±0.014 \\
Education Training  & -0.032±0.015 & 0.016±0.014 \\
Finance  & 0.000 & -0.044±0.020 \\
Government Public Administration  & 0.000 & -0.009±0.017 \\
Health Science  & 0.000 & -0.026±0.013 \\
Hospitality Tourism  & 0.000 & 0.000 \\
Human Services  & 0.024±0.016 & 0.000 \\
Information Technology  & 0.000 & 0.000 \\
Law Public Safety Corrections Security  & 0.000 & -0.018±0.020 \\
Manufacturing  & -0.001±0.012 & -0.017±0.013 \\
Marketing  & 0.000 & -0.009±0.022 \\
Science Technology Engineering Mathematics  & 0.000 & -0.022±0.014 \\
Transportation Distribution Logistics  & 0.000 & -0.014±0.017 \\
        \bottomrule
    \end{tabular}}
    \caption{Effects of of \Resume{} Language (Sentiment Analysis) factors on predicting seniority. Reported coefficients in logistic regression with standard error intervals. Larger positive numbers imply a greater positive impact for senior transitions, while more negative numbers imply a penalty, values near zero imply little effect. }
    \label{tab:lang_factors_senitment}
\end{table}

\begin{table*}[ht]
    \centering
    \resizebox{1.0\textwidth}{!}{%
    \begin{tabular}{l|rrrr}
        \toprule
        \textbf{Job Sector} & 
\textbf{ Hedges } &
\textbf{ Graded Quantifiers} &
\textbf{ Qualifiers } &
\textbf{ Buzzwords } \\
\midrule
Agriculture Food Natural Resources  & 0.000 & -0.018±0.019 & 0.000 & 0.155±0.017 \\
Architecture Construction  & 0.000 & 0.000 & 0.000 & 0.198±0.015 \\
Arts Audio Video Technology Communications  & 0.000 & 0.000 & 0.000 & 0.226±0.016 \\
Business Management Administration  & 0.000 & -0.005±0.014 & 0.000 & 0.250±0.014 \\
Education Training  & 0.002±0.015 & 0.000 & 0.007±0.013 & 0.154±0.014 \\
Finance  & -0.033±0.019 & 0.000 & 0.000 & 0.197±0.019 \\
Government Public Administration  & 0.000 & 0.000 & -0.014±0.016 & 0.200±0.017 \\
Health Science  & 0.016±0.013 & 0.000 & 0.000 & 0.173±0.012 \\
Hospitality Tourism  & 0.000 & 0.000 & -0.027±0.019 & 0.071±0.018 \\
Human Services  & 0.000 & 0.000 & 0.000 & 0.086±0.016 \\
Information Technology  & 0.000 & 0.000 & 0.000 & 0.239±0.033 \\
Law Public Safety Corrections Security  & 0.000 & 0.000 & 0.000 & 0.156±0.020 \\
Manufacturing  & -0.020±0.013 & 0.000 & -0.004±0.012 & 0.291±0.013 \\
Marketing  & 0.000 & 0.000 & 0.000 & 0.224±0.023 \\
Science Technology Engineering Mathematics  & 0.000 & 0.000 & -0.024±0.013 & 0.221±0.015 \\
Transportation Distribution Logistics  & 0.000 & 0.000 & 0.000 & 0.277±0.016 \\
        \bottomrule
    \end{tabular}}
    \caption{Effects of of \Resume{} Language (Phrases) factors on predicting seniority. Reported coefficients in logistic regression with standard error intervals. Larger positive numbers imply a greater positive impact for senior transitions, while more negative numbers imply a penalty, values near zero imply little effect.}
    \label{tab:lang_factors_phrases}
\end{table*}

\begin{table*}[ht]
    \centering
    \resizebox{1.0\textwidth}{!}{%
    \begin{tabular}{l|rrrrrrrr}
        \toprule
        \textbf{Job Sector} & 
\textbf{ Passive Voice } &
\textbf{ Fixed Dependencies } &
\textbf{ Vocative Dependencies} &
\textbf{ Expletive Dependencies } &
\textbf{ Discourse Dependencies} &
\textbf{ Past Tense } &
\textbf{ Pronouns } &
\textbf{ "To Be" Words } \\ \midrule
Agriculture Food Natural Resources  & 0.000 & 0.000 & 0.000 & -0.024±0.019 & 0.000 & 0.000 & -0.037±0.020 & 0.000 \\
Architecture Construction  & 0.000 & 0.000 & 0.000 & 0.000 & 0.000 & 0.000 & -0.045±0.017 & 0.000 \\
Arts Audio Video Technology Communications  & 0.000 & 0.000 & 0.000 & 0.000 & 0.000 & 0.000 & -0.052±0.017 & 0.000 \\
Business Management Administration  & 0.000 & 0.000 & 0.000 & -0.029±0.014 & 0.000 & 0.000 & -0.067±0.015 & 0.000 \\
Education Training  & 0.000 & 0.000 & 0.000 & -0.004±0.014 & 0.000 & 0.028±0.017 & 0.000 & -0.047±0.019 \\
Finance  & 0.000 & 0.000 & 0.000 & 0.000 & 0.000 & 0.000 & -0.022±0.020 & 0.000 \\
Government Public Administration  & 0.000 & 0.000 & 0.000 & 0.000 & 0.000 & 0.000 & -0.044±0.016 & 0.000 \\
Health Science  & 0.020±0.012 & 0.000 & 0.000 & -0.003±0.012 & 0.000 & 0.046±0.014 & -0.052±0.014 & 0.000 \\
Hospitality Tourism  & 0.000 & 0.000 & 0.000 & 0.000 & 0.000 & 0.000 & 0.000 & 0.000 \\
Human Services  & 0.000 & 0.000 & 0.000 & 0.000 & 0.000 & 0.000 & 0.000 & 0.000 \\
Information Technology  & 0.000 & 0.000 & 0.000 & 0.000 & 0.000 & 0.000 & 0.000 & 0.000 \\
Law Public Safety Corrections Security  & 0.000 & 0.000 & 0.000 & 0.000 & 0.000 & 0.007±0.020 & 0.000 & 0.000 \\
Manufacturing  & 0.005±0.012 & 0.000 & 0.000 & -0.020±0.013 & 0.000 & 0.007±0.015 & -0.063±0.014 & 0.000 \\
Marketing  & 0.000 & 0.000 & 0.000 & 0.000 & 0.000 & 0.000 & 0.000 & 0.000 \\
Science Technology Engineering Mathematics  & 0.000 & 0.000 & 0.000 & 0.000 & 0.000 & 0.053±0.016 & -0.062±0.014 & 0.000 \\
Transportation Distribution Logistics  & 0.001±0.022 & 0.000 & 0.000 & 0.000 & 0.000 & 0.000 & -0.064±0.018 & 0.000 \\
        \bottomrule
    \end{tabular}}
    \caption{Effects of of \Resume{} Language (Syntax) factors on predicting seniority. Reported coefficients in logistic regression with standard error intervals. Larger positive numbers imply a greater positive impact for senior transitions, while more negative numbers imply a penalty, values near zero imply little effect.}
    \label{tab:lang_factors_syntax}
\end{table*}

\bibliographystyle{ACM-Reference-Format}
\bibliography{references}

\end{document}